\documentclass[12pt]{article}

\usepackage{epsfig}
\usepackage{cite}
\usepackage{amsmath, amssymb, amsfonts}
\usepackage{color}
\usepackage{latexsym}
\usepackage{graphicx}
\usepackage[colorlinks,bookmarks]{hyperref}
\hypersetup{pdfpagemode=UseNone, pdfstartview=FitH, linkcolor=blue,
            citecolor=red, urlcolor=blue}

\def\be{\begin{equation}}
\def\ee{\end{equation}}
\def\ba{\begin{eqnarray}}
\def\ea{\end{eqnarray}}

\def\bdm{\begin{displaymath}}
\def\edm{\end{displaymath}}
\def\la{~\mbox{\raisebox{-.6ex}{$\stackrel{<}{\sim}$}}~}
\def\ga{~\mbox{\raisebox{-.6ex}{$\stackrel{>}{\sim}$}}~}
\def\bq{\begin{quote}}
\def\eq{\end{quote}}

 at 10truept
 at 14truept

\newcommand{\p}{\partial}

\newcommand{\Mpl}{M_{\mathrm{Pl}}}
\newcommand{\mps}{M_{\mathrm{Pl}}^2}

\newcommand{\bea}{\begin{eqnarray}}
\newcommand{\eea}{\end{eqnarray}}

\newcommand{\bi}{\begin{itemize}}
\newcommand{\ei}{\end{itemize}}

\newcommand{\beq}{\begin{equation}}
\newcommand{\eeq}{\end{equation}}
\newcommand{\beqa}{\begin{eqnarray}}
\newcommand{\eeqa}{\end{eqnarray}}
\newcommand{\mpl}{\Mpl}

\def\la{~\mbox{\raisebox{-.6ex}{$\stackrel{<}{\sim}$}}~}
\def\ga{~\mbox{\raisebox{-.6ex}{$\stackrel{>}{\sim}$}}~}

\def\ltap{\ \raise.3ex\hbox{$<$\kern-.75em\lower1ex\hbox{$\sim$}}\ }
\def\gtap{\ \raise.3ex\hbox{$>$\kern-.75em\lower1ex\hbox{$\sim$}}\ }
\def\gl{\ \raise.5ex\hbox{$>$}\kern-.8em\lower.5ex\hbox{$<$}\ }
\def\roughly#1{\raise.3ex\hbox{$#1$\kern-.75em\lower1ex\hbox{$\sim$}}}

\begin{document}

\thispagestyle{empty}
\begin{flushright}
July 2023 \\
\end{flushright}
\vspace*{1.75cm}
\begin{center}

{\Large \bf Axion Flux Monodromy Discharges Relax}
\vskip.3cm 
{\Large \bf the Cosmological Constant} 

\vspace*{1.25cm} {\large 
Nemanja Kaloper$^{a, }$\footnote{\tt
kaloper@physics.ucdavis.edu} 
}\\
\vspace{.2cm} 
{\em $^a$QMAP, Department of Physics and Astronomy, University of
California}\\
\vspace{.05cm}{\em Davis, CA 95616, USA}\\

\vspace{1.7cm} ABSTRACT
\end{center}
Linear axion monodromy models modulated with higher powers of fields naturally realize the quantum-mechanical flux discharge 
mechanism for relaxing the cosmological constant toward zero. Working with multiple copies of superposed linear and quadratic 
flux monodromies, each copy spanned by a pair of fluxes, we show that when the axion is very massive and so effectively decoupled, 
the membrane discharges relax the cosmological constant toward an attractor $0 < \Lambda/\mpl^4 \ll 1$. If we restrict the flux 
variations and the intermediate flux values to never venture beyond a finite flux range, the terminal value of the cosmological 
constant will be tiny but finite. We show how it can reproduce the observed scale of dark energy, and explain how to incorporate 
matter sector phase transitions.

\vfill \setcounter{page}{0} \setcounter{footnote}{0}

\vspace{1cm}
\newpage

\section{Introduction}

In a recent series of papers \cite{Kaloper:2022oqv,Kaloper:2022utc,Kaloper:2022jpv,Kaloper:2023xfl} we have been formulating 
a novel approach to the cosmological constant problem, viewing the cosmological constant as an environmental variable, controlled 
by the fluxes of non-propagating $4$-form field strengths. In the presence of charged membranes, these fluxes change discretely
\cite{Brown:1987dd,Brown:1988kg} and lead to a multiverse of bubbles which scan $\Lambda$. An operational framework
which leads to a sufficiently refined spectrum of universes and favors the almost-Minkowski space as the unique attractor is well 
approximated by the covariant unimodular gravity framework of Henneaux and Teitelboim \cite{Henneaux:1989zc}, extended 
with the inclusion of charged tensional membranes. Any de Sitter space decays by nucleation of bubbles bounded 
by the membranes which reduce the $4$-form flux.  To ensure the states are very finely grained, coming very close to 
vanishing cosmological constant, we used at least two $4$-forms and their associated membrane towers with fluxes which 
are mutually incommensurate. Then when the flux energy is dominated by linear terms in $4$-form fluxes, an almost-Minkowski 
space is the unique long-time attractor. A very small cosmological constant is natural in all such frameworks, 
without invoking anthropic reasoning.

Questions naturally arise about where the linear flux terms come from, must they be mutually incommensurate, and can they be UV completed in 
some controllable regime of, e.g., field theory. The explicit realizations of the relaxation mechanisms 
we explored in \cite{Kaloper:2022oqv,Kaloper:2022utc,Kaloper:2022jpv,Kaloper:2023xfl} employ the 
same ingredients as the landscapes of axion monodromy, encountered in the construction of models of inflation and dark energy 
\cite{Silverstein:2008sg,McAllister:2008hb,Kaloper:2008qs,Kaloper:2008fb,Flauger:2009ab,Dong:2010in,Kaloper:2011jz,Kaloper:2014zba,Marchesano:2014mla,McAllister:2014mpa,Kaloper:2016fbr,Montero:2017yja,DAmico:2017cda,Buratti:2018xjt}. 
This raises the possibility that our cosmological constant relaxation by flux discharge can be realized in monodromy field theories. 

In this work we confirm this and give a realization of the adjustment mechanism with monodromies. 
A clue to how to design the mechanism comes from considering the problems of the old Abbott adjustment mechanism \cite{Abbott:1984qf}, 
which we compared with the flux adjustment earlier  \cite{Kaloper:2022oqv,Kaloper:2022utc,Kaloper:2022jpv,Kaloper:2023xfl}. The Abbott mechanism
involves a scalar field with a linear potential modulated with a periodic function,
\be
V(\phi) = - \alpha \mu^3 \phi + m^4 \bigl(1-\cos(\frac{\phi}{f}) \bigr) \, ,
\label{abbott}
\ee
which by universality of gravity is degenerate with the cosmological constant $\Lambda$, giving the total vacuum energy
$\frac14 \langle 0 | T^\mu{}_\mu | 0 \rangle = \Lambda + V(\phi)$.  Here $\alpha$ is a dimensionless number, $\mu$ a mass scale, 
and $m$ the gauge theory strong coupling scale suppressed by the instanton dilute gas action, 
$m^4 \simeq M_{\tt QFT}^4 \, e^{-S_{inst}}$. 
The idea is that $\frac14 \langle 0 | T^\mu{}_\mu  | 0 \rangle$ is 
gradually adjusted to zero by the \underline{classical slow roll} of the scalar down the potential slope, which occurs since $\mu$ and $m$ are small.
Only when $\phi$ approaches the critical value where $\frac14 \langle 0 | T^\mu{}_\mu  | 0 \rangle \simeq 0$ do the barriers in the
potential manage to catch and stop the scalar. The scalar rolling must be slow because if the 
cosmological constant remainder after the scalar is trapped by a well is to be sufficiently small, the barriers must be very subtle. So 
unless the linear term is shallow, the scalar won't be caught before the universe stops to expand and collapses. On the other hand
this also yields the empty universe problem \cite{Abbott:1984qf}. The scalar in slow roll always dominates the expansion, and so
when it finally stops, there is not enough energy to reheat anything except photons and, maybe, neutrinos. 

Hence the root cause of the empty universe problem of the Abbott proposal  \cite{Abbott:1984qf} is that the field in slow roll is migrating 
from one state of the universe to another in very tiny steps. The fact that this is a smooth process is not so important, as one 
can see from the example of discrete variation of the cosmological constant in tiny steps of $\simeq 10^{-120} \mpl^4$, where the 
same problem reoccurs \cite{Brown:1987dd,Brown:1988kg}. In either case, the universe remains dominated by the effective vacuum energy, and so 
once it finally decays, there is not much left of it. A resolution was pointed out by \cite{Bousso:2000xa}, whereby the cosmological
constant changes in very large jumps from one stage to another, and the tiny terminal value is achieved by a ``misalignment" of
the successive discharges that keep reducing it. We have provided a different specific mechanism for achieving a small terminal value
of $\Lambda$ in \cite{Kaloper:2022oqv,Kaloper:2022utc,Kaloper:2022jpv,Kaloper:2023xfl}. 

Here we show how to realize a variant of our mechanism of relaxation of the cosmological constant 
in field theory. Since we are using axion monodromies to span the spectrum of values of the cosmological constants, which
yield natural screening, we should restrict the field and flux ranges. This implies that the resulting discretuum will be grainy, 
and that the terminal value of $\Lambda$ is gapped from zero. To realize the required density of the spectrum, which is able to reproduce the
gap $\Delta \Lambda \la 10^{-120} \mpl^4$, we must use at least ${\cal O}(20)$ distinct sectors, instead of just two as in 
\cite{Kaloper:2022oqv,Kaloper:2022utc,Kaloper:2022jpv,Kaloper:2023xfl}. However the adjustment of $\Lambda$ by screening and
discharge remains natural, without invoking anthropics. 

\section{Model Building}

We start by noting that the potential (\ref{abbott}) is very similar
in form to the leading order contributions in the relaxion scenario of \cite{Graham:2015cka}, but below the gauge theory strong coupling scale, 
such that the instanton-induced potential wells and barriers are not negligible around $V \approx 0$. This would have been the problem for
the relaxion scenario, which exploits the scalar's evolution in the slow roll, just like Abbott's model. Here, however, to circumvent the
empty universe problem, we wish to block the classical regime, and the large potential barriers are not only welcome but required. 

Note also, that we wish to ensure 
that the potential (\ref{abbott}) remains a valid description of evolution over a large range of scales toward zero.
As noted in \cite{Ibanez:2015fcv,McAllister:2016vzi}, a tool available to ensure this is to 
complete (\ref{abbott}) into a monodromy structure, where $V(\phi)$ displayed in (\ref{abbott}) is but a single branch. With replacing 
$\alpha \mu^3\phi \rightarrow \alpha \mu^2 (\mu \phi +Q)$, where $Q$ is the magnetic dual of a massless $4$-form $F_{\mu\nu\lambda\sigma}$,
in the regime where flattening induces dominant linear term in the potential, (\ref{abbott}) is 
replaced by\footnote{To allay disconcert of an attentive reader about adding fluxes $Q$, about 
where the field strengths are, 
we remind that the $Q$-dependent terms in (\ref{abbottmon}) come 
from dualizing 
$-\frac{1}{48} F_{\mu\nu\lambda\sigma}^2  - \frac{\mu(\alpha \mu + \phi)}{24} F_{\mu\nu\lambda\sigma} \epsilon^{\mu\nu\lambda\sigma}/\sqrt{g}$
and so on. Many more details can be found in \cite{Kaloper:2022oqv,Kaloper:2022utc,Kaloper:2022jpv,Kaloper:2023xfl,Dvali:2005an,Kaloper:2008qs,Kaloper:2008fb,Kaloper:2016fbr}. 
We will use this shortcut throughout this work.}
\be
V(\phi,Q) = - \alpha \mu^2 \bigl(\mu \phi +Q \bigr)  + \frac{\zeta}{2} \bigl(\mu \phi+Q \bigr)^2 + m^4 \bigl(1-\cos(\frac{\phi}{f} ) \bigr) + \ldots \, ,
\label{abbottmon}
\ee
where the magnetic flux $Q$ is quantized, $Q = N q$ (see, e.g. 
references \cite{Teitelboim:1985ya,Teitelboim:1985yc}). We retained the quadratic correction to Abbott's linear term, just like in the relaxion scenario,  
although it will play a more prominent role for us, to be explicitly elaborated below. We also introduced the ellipsis to 
denote higher order corrections to the linear potential expected to arise in realistic constructions. 

We choose the quantization conditions that justify \underline{not} including 
$Q$ in the cosine, as follows. 
The unit of charge $q$ is the membrane charge which sources $Q$, and we must take 
\be
 2\pi \mu f = \frac{r}{s} q = q \, , 
 \label{quant}
\ee
where in general $r$ and $s$ are two prime integers\footnote{If $r/s$ were an irrational number the field $\phi$ would decompactify, just like the
irrational axion of \cite{Banks:1991mb}. This would be in conflict with the lore that quantum gravity 
does not tolerate global symmetries \cite{Banks:2010zn}.}. 
This ensures that the field $\phi$ has the range $0 \le \phi < 2\pi f$, which is compact, so that 
interpreting it as a monodromy axion is consistent \cite{Silverstein:2008sg,McAllister:2008hb,Kaloper:2008qs,Kaloper:2008fb}. 
Here we take $r/s=1$ for simplicity. As a result the 
axion shift symmetries remain discrete, and specifically $\phi \rightarrow \phi + 2\pi f$ and $Q \rightarrow Q - q$ is a symmetry of (\ref{abbottmon}). 
As a consequence, when $r/s = 1$, the phase in the cosine after the shift 
$\alpha \mu^3\phi \rightarrow \alpha \mu^2 (\mu \phi +Q)$ would be 
$\frac{Q}{\mu f} = N \frac{q}{\mu f} = 2\pi N$, and so $Q$ drops out of the cosine
term in Eq. (\ref{abbottmon}). The result is a flattened monodromy model with a 
cosine modulation, which has been studied in 
cosmology \cite{Flauger:2009ab,Dong:2010in,McAllister:2014mpa}, in slow roll regime of $\phi$. 
Given the protection mechanisms\footnote{Including the discrete shift symmetries and the continuous gauge symmetry 
of the dual electric theory, whose role is discussed in detail in \cite{Kaloper:2008fb,Kaloper:2011jz,Kaloper:2016fbr,DAmico:2017cda}.} 
of monodromy constructions, the range of validity of $V$ in (\ref{abbottmon}) is all the way up to some large cutoff ${\cal M} \la \mpl$. If
the mass $\mu$ is small, then $\phi$ can also vary over transplanckian ranges \cite{Silverstein:2008sg,McAllister:2008hb,Kaloper:2008qs,Kaloper:2008fb,Flauger:2009ab,Dong:2010in,Kaloper:2011jz,Kaloper:2014zba,Marchesano:2014mla,McAllister:2014mpa,Kaloper:2016fbr,Montero:2017yja,DAmico:2017cda,Buratti:2018xjt}. 
However, we will not need such large $\phi$ variations. We will return to this issue shortly. 

Note that both harmonic terms in (\ref{abbottmon}) are in phase. This actually is not a restriction because of our 
quadratic truncation of the monodromy spanning terms. If we allow for a phase difference between the two harmonic 
functions in (\ref{abbottmon}), and encode it in the monodromy sector by replacing the relevant terms in (\ref{abbottmon}) 
by $-\alpha \mu^2 (\mu(\phi-\phi_0) +Q) + \zeta (\mu(\phi-\phi_0) +Q)^2/2$, and then expand this quadratic
in terms of the powers of $\mu \phi + Q$, we get
\be
\frac{\zeta}{2} \mu^2 \phi_0^2  - (\alpha \mu^2 + \zeta \mu \phi_0) \bigl(\mu \phi +Q \bigr)  + \frac{\zeta}{2} \bigl(\mu \phi +Q \bigr)^2 \, .
\ee
The leading constant can be absorbed into the cosmological constant (to be included below), and the coefficient 
of the linear monodromy branch terms can be redefined by $\alpha \mu^2 + \zeta \mu \phi_0 \rightarrow \alpha \mu^2$ 
to restore the form of (\ref{abbottmon}). Therefore we will ignore this phase difference in
what follows. 

\begin{figure}[bth]
    \centering
    \includegraphics[width=12cm]{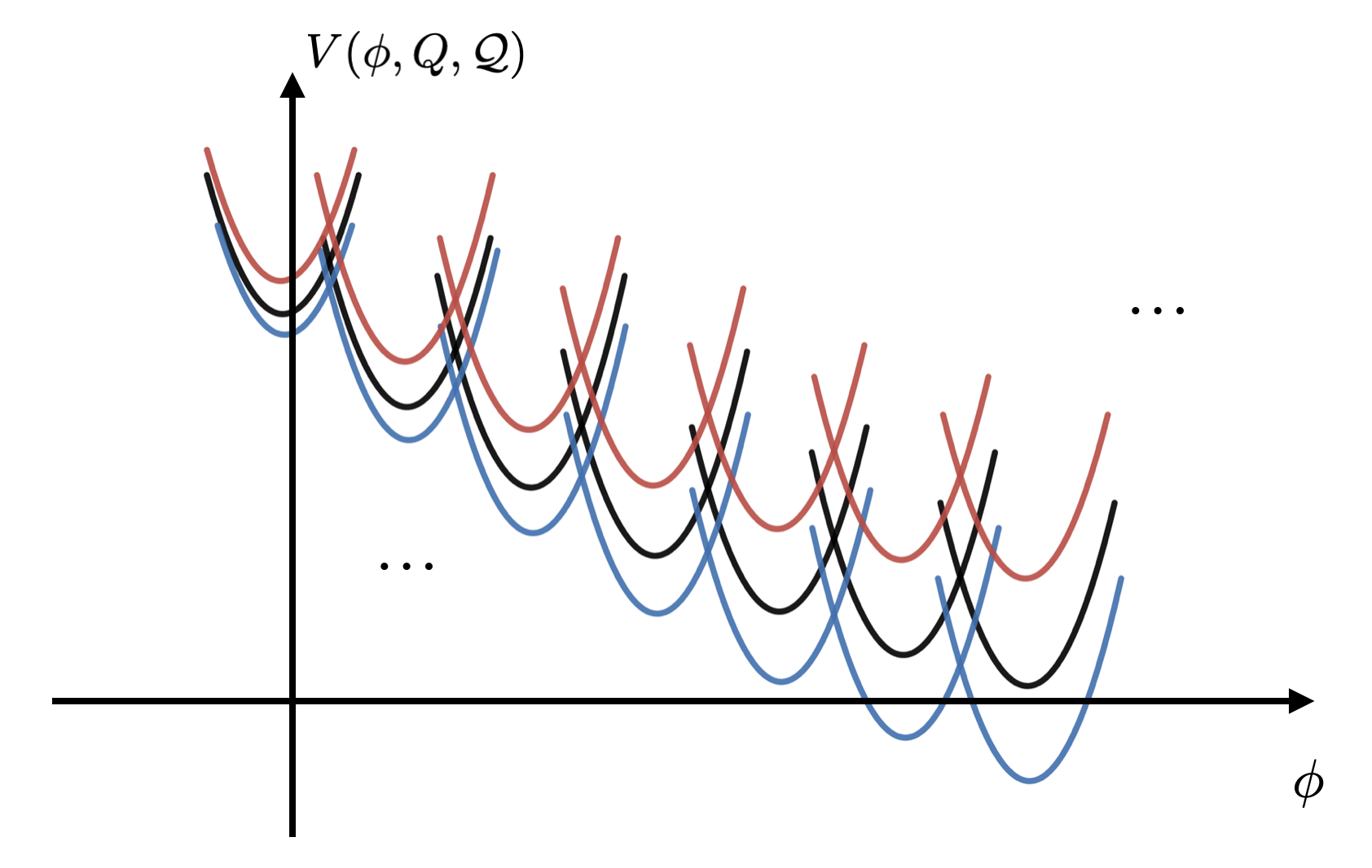}
    \caption{Double monodromy potential of Eq. (\ref{abbottfin}). The same color denotes the same $Q = {\rm const}$ branch.
    The segments of parabola denote the parabolic branches along which ${\cal Q}$ is fixed. 
    The other branches to the left and to the right are indicated by ellipses.}
    \label{figpot}
\end{figure}

To simplify our model further, we recall that the instanton dilute gas approximation which yields the cosine modulation in 
(\ref{abbott}) and (\ref{abbottmon}) relies on weak gauge couplings and large instanton actions. As the couplings become stronger and
instanton actions decrease, there are additional harmonic corrections to Eqs. (\ref{abbott}), (\ref{abbottmon}). 
In the large ${\cal N}$ limit this series can be reorganized and one finds that axion potentials 
develop additional monodromy branches \cite{Witten:1980sp,DiVecchia:1980yfw,Ohta:1981ai,Dubovsky:2011tu,Lawrence:2012ua}. 
The simplest form of the potential are the quadratics discussed
in \cite{Kaloper:2008qs,Kaloper:2008fb}, but more general potentials can also appear \cite{Dvali:2005an,Kaloper:2014zba,Kaloper:2016fbr,DAmico:2017cda}. It will suffice here to stop at the quadratic limit 
of \cite{Kaloper:2008qs,Kaloper:2008fb}. We replace the cosine terms in (\ref{abbottmon}) with a quadratic monodromy, 
spanned by the magnetic flux ${\cal Q}$,
\be
V(\phi,Q,{\cal Q}) = - \alpha \mu^2 \bigl(\mu \phi +Q \bigr)  + \frac{\zeta}{2} \bigl(\mu \phi +Q \bigr)^2 
+ \frac{\beta}{2 } \bigl(\mu \phi + {\cal Q} \bigr)^2 + \ldots \, ,
\label{abbottfin}
\ee
where $\beta = m^4/\mu^4$, ${\cal Q}$ obeys the same quantization rule  (\ref{quant}) 
as $Q$, reflecting our choice of the period $2\pi f$ above, 
and ellipsis again denote higher order corrections. We plot this potential in
Fig. (\ref{figpot}). Basically, what it is, functionally, is the potential (\ref{abbottmon}) 
where the cosine is expanded to quadratic order, and then the 
resulting parabola is extended ad infinitum, while periodicity is restored by adding 
${\cal Q}$, which includes infinitely many parabolas to each linear
branch, that is getting progressively more distorted at large field values. 

In the classical limit, the parabolic sections of the potential (\ref{abbottfin}) are completely separated from each other. The field $\phi$, once
deposited into one of them, can never change it by classical roll. Quantum corrections could re-link them by mixing and level repulsion
of degenerate states, producing a system of allowed and forbidden energy bands \cite{Kaloper:2011jz}. 
However that will be of little consequence here. Unlike our original use of flux monodromy to support slow roll inflation 
or quintessence \cite{Kaloper:2008qs,Kaloper:2008fb,Kaloper:2011jz}, or in the relaxion scenario, 
\cite{Graham:2015cka}, which are designed to operate entirely in the classical roll regime until extremely 
low scales, we will completely freeze out classical motion of $\phi$. This means, we set $\phi$ into one of the minima of (\ref{abbottfin}), 
defined by $\partial_\phi V(\phi,Q,{\cal Q}) = - \alpha \mu^3 + \zeta \mu \bigl(\mu\phi  + Q \bigr) + \beta \mu \bigl(\mu \phi  + {\cal Q} \bigr)  = 0$,
\be
\mu \phi_{min}  = \frac{\alpha \mu^2 - \beta {\cal Q} - {\zeta}Q}{\beta+\zeta}  \, ,
\label{minpot}
\ee
and to make sure it stays put there, we require that 
\be
m_\phi^2 = \partial^2_\phi V = (\beta +\zeta) \mu^2 \gg H^2 = \frac{V + \Lambda}{3\mpl^2} \, ,
\label{mass}
\ee
where we model the contribution of all other sectors of the universe to the stress energy tensor by their cosmological 
constant $\Lambda$. This is reasonable, since as long as the cosmological constant $\Lambda$ is large, it will rapidly dilute
all non-constant contributions to $T^\mu{}_\nu$. This immediately shows that the regime we work in is exactly the opposite of the
domain where relaxion operates \cite{Graham:2015cka}, as it does not satisfy neither of the ``three commandments" of relaxion dynamics: 
\begin{itemize}
\item[] 1) the evolution is vacuum energy dominated,
\item[] 2) the barriers to the $\phi$ roll are already present, being generated at some very high scale ${\cal M}$ which serves as
the UV cutoff of the theory, 
\item[] 3) the barriers are always steep enough to catch and hold $\phi$ since the quadratic monodromy 
branches are mutually disjoint; the total potential (\ref{abbottfin}) is never flat enough for $\phi$ to roll off. 
\end{itemize}

Note that $\phi$ mass does not 
depend on the value of $\phi$, $Q$ or ${\cal Q}$. Once (\ref{mass}) holds for some $V + \Lambda$, it is {\it guaranteed} to hold for
all smaller values of $V + \Lambda$.
Thus we can integrate $\phi$ out by substituting (\ref{minpot}) into (\ref{abbottfin}), to find  
\be
V(\phi_{min},Q,{\cal Q}) = -\frac{\alpha^2 \mu^4}{2(\beta+\zeta)} +\frac{\beta}{\beta+\zeta} \alpha \mu^2 \bigl({\cal Q} - Q \bigr) 
+\frac{\beta\zeta}{2(\beta+\zeta)} \bigl({\cal Q} - Q \bigr)^2  + 
\ldots \, .
\label{abbottfinz}
\ee
The potential depends only on the difference of fluxes ${\cal Q} -Q$, in order to maintain unbroken 
diagonal discrete shift symmetry $Q \rightarrow Q+ nq$, 
${\cal Q} \rightarrow {\cal Q} + nq$. While $Q$ and ${\cal Q}$ seem to be completely degenerate, 
they are independent discrete degrees of freedom. This could be seen if we take different (mutually rational) 
charges q for each of them, and so we retain both of them explicitly.

Although $\phi$ is fixed, the potential (\ref{abbottfinz}) can still 
change\footnote{There are corrections due to the quantum dynamics of $\phi$ which we ignore here. 
First, in de Sitter space, quantum fluctuations ``smear" $\phi$ by 
$\delta \phi \la H^2/m_\phi$ \cite{Linde:1990flp}. Secondly, there can be processes where the field 
tunnels through from one parabolic branch to another by keeping the same value of ${\cal Q}$. In this case 
the field will end up displaced from the minimum in the new branch, and immediately after the  bubble 
forms the field will start to oscillate, dissipating the extra unit of ${\cal Q}$. Since the bubble interior is still 
controlled by the cosmological constant and the field mass is large, the oscillations will dissipate away quickly and this process will be more 
suppressed than the direct discharge of ${\cal Q}$.} by bubble nucleation induced by flux discharge of $Q$ and ${\cal Q}$.
We can view this as a thin bubble wall limit of tunneling in field theory \cite{Coleman:1980aw,Parke:1982pm}, in line 
with our replacement of the instanton-generated cosine potential of Eq. (\ref{abbottmon}) with monodromy branches (\ref{abbottfin}). 
We model those tunneling processes by simply including 
tensional charged membranes which source $Q$ and ${\cal Q}$. Such a limit could be realized by considering the
strongly coupled gauge theories in the large ${\cal N}$ limit exhibited in \cite{Dubovsky:2011tu,Lawrence:2012ua,Nomura:2017ehb}. Here
we will simply add such membranes and treat their charges and tensions as free parameters, which 
can be calculated in principle from 
a microscopic realization of the monodromy model. We will check however that at least nominally the membranes may obey the bounds
arising from the Weak Gravity Conjecture \cite{Arkani-Hamed:2006emk}, 
although since they are emergent low energy structures that may not even be necessary \cite{Saraswat:2016eaz}. 

Therefore, following \cite{Kaloper:2022oqv,Kaloper:2022utc,Kaloper:2022jpv,Kaloper:2023xfl},
(and adopting the normalization ``translation" from the conventions of \cite{Kaloper:2022oqv,Kaloper:2022utc,Kaloper:2022jpv,Kaloper:2023xfl} to here 
$\lambda = Q/2$, $\hat \lambda = {\cal Q}/2$) we extend the effective action of the axion monodromy with dynamical boundary terms\footnote{Which, again,
arise from the fact that $Q$ and ${\cal Q}$ are a part of magnetic duals of $4$-form field strengths as noted above.}
\ba
S_{membranes} &=& 
\int d^4x \Bigl\{\frac{1}{6} {\epsilon^{\mu\nu\lambda\sigma}} (\partial_\mu Q ) {\cal A}_{\nu\lambda\sigma} 
+ \frac{1}{6} {\epsilon^{\mu\nu\lambda\sigma}} (\partial_\mu {\cal Q}) {\cal A}'_{\nu\lambda\sigma}  \Bigr \} 
 \nonumber \\
&&~~~~~~  - {\cal T}_{Q} \int d^3 \xi \sqrt{\gamma} - q \int {\cal A} - {\cal T}_{\cal Q} \int d^3 \xi \sqrt{\gamma} 
- q \int {\cal A}' \, . 
\label{actionnewmemdcharg} 
\ea
Here ${\cal T}_{i}$ and $q$ are the membrane tensions and charges, respectively. Recall that $Q$ and ${\cal Q}$ 
are sourced by membranes which have equal unit charges $q$ to ensure the axion $\phi$ has finite period. 
Since we are working in the dual magnetic picture of $4$-forms,
${\cal A}, {\cal A}'$ play the role of auxiliary fields locally enforcing $\partial_\mu Q = \partial_\mu {\cal Q}=  0$.  
We also need to add  
the Israel-Gibbons-Hawking term $-  \int d^3 \xi \sqrt{\gamma} \mpl^2  \, \bigl[ K \bigr]$ 
for gravity which encodes boundary conditions across membrane walls, and $\bigl[K\bigr] = K_+ - K_-$ 
is the jump across a membrane. The topological terms 
$\propto {\epsilon^{\mu\nu\lambda\sigma}} (\partial_\mu Q) {\cal A}_{\nu\lambda\sigma}$ etc do not 
gravitate since they don't depend on the metric. They
fix $Q, {\cal Q}$ to be locally constant, changing only across membrane walls. 
The charge terms are 
\be
\int {\cal A} = \frac16 \int d^3 \xi {\cal A}_{\mu\nu\lambda} \frac{\p x^\mu}{\p \xi^\alpha} \frac{\p x^\nu}{\p \xi^\beta} 
\frac{\p x^\lambda}{\p \xi^\gamma} \epsilon^{\alpha\beta\gamma} \, .
\ee
We take ${\cal T}_{i} > 0$ to prevent any problems with instabilities and ghosts. Note, that in principle the same
calculations as we are about to perform could be done with harmonic potentials, working out tunneling rates
between adjacent vacua. Membrane nucleations are merely a thin wall limit of such processes. 

\section{Discharging $\Lambda$: the Attractor Regime}

To study quantum membrane discharge in the semiclassical limit we Wick-rotate the full theory 
to Euclidean time. We determine the instanton configurations which mediate transitions catalyzed by membranes 
and calculate the rates $\Gamma \sim e^{-S_E}$, where $S_E$ is the Euclidean bounce action \cite{Coleman:1980aw}.  
All the relevant details are presented in  \cite{Kaloper:2022oqv,Kaloper:2022utc,Kaloper:2022jpv,Kaloper:2023xfl}, and we will not
repeat all of the derivations here, but direct an interested reader to those references. We will only 
recapitulate the salient features of the argument and the answer here.

Since we only consider transitions between  
locally maximally symmetric backgrounds with local $O(4)$ symmetry, which are the dominant processes, 
we can write the total cosmological constant in any patch $\Lambda_{\tt total} = \Lambda_{\tt QFT} + V(\phi_{min},Q,{\cal Q})$ as
(setting aside higher order corrections for the time being)
\ba
\Lambda_{\tt total} &=&  \Lambda_{\tt QFT} -\frac{\alpha^2 \mu^4}{2(\beta+\zeta)} +\frac{\beta}{\beta+\zeta} \alpha \mu^2 \bigl({\cal Q} - Q \bigr) 
+\frac{\beta\zeta}{2(\beta+\zeta)} \bigl({\cal Q} - Q \bigr)^2  \nonumber \\
&=&  \Lambda_0 +\frac{\beta}{\beta+\zeta} \alpha \mu^2 \bigl({\cal Q} - Q \bigr) 
+\frac{\beta\zeta}{2(\beta+\zeta)} \bigl({\cal Q} - Q \bigr)^2 \, ,
\label{totalcc}
\ea
where $Q, {\cal Q}$ can vary from patch to patch across membrane walls. 
We absorbed  $-\frac{\alpha^2 \mu^4}{2(\beta+\zeta)}$ into $\Lambda_0$. 

To describe these transitions, 
we take the local Euclidean geometry to be patches of $S^4$, 
with the metrics $ds^2_E =  dr^2 + a^2(r) \, d\Omega_3$, and glue together two sections with different curvatures 
along a latitude line, with the junction conditions on the latitude enforcing the curvature discontinuity and the location of the latitude, 
\be
a_{out} = a_{in} \, , ~~~~~~~~   \Delta Q = \Delta{\cal Q} =  q \, ,  ~~~~~~~~ \mps \Bigl(\frac{a_{out}'}{a} - \frac{a_{in}'}{a} \Bigr)
 = -\frac{{\cal T}_{j}}{2} \, ,
 \label{metricjc}
\ee 
with $j = Q$ or ${\cal Q}$, bearing in mind that only one occurs at a time. 
 
Boundary conditions (\ref{metricjc}) combined with bulk equations away from a membrane can be solved to yield 
the junction conditions on the membrane. Straightforward calculation 
yields \cite{Kaloper:2022oqv,Kaloper:2022utc,Kaloper:2022jpv,Kaloper:2023xfl}, using $\Delta \Lambda = \Lambda_{out} - \Lambda_{in}$,
\ba
\zeta_{out} \sqrt{ 1-  \frac{\Lambda_{out} a^2}{3 \mps}} 
&=& - \frac{{\cal T}_{j}}{4\mps}\Bigl(1 -  \frac{4\mps}{3 {\cal T}_{j} } \Delta \Lambda \Bigr)\, a \, , \nonumber \\
\zeta_{in} \sqrt{1-  \frac{\Lambda_{in} a^2}{3 \mps}} 
&=& \frac{{\cal T}_{j}}{4\mps}\Bigl(1 +  \frac{4\mps}{3 {\cal T}_{j} }\Delta \Lambda \Bigr) \, a \, . 
\label{diffroots}
\ea
Here $\zeta_j = \pm$ picks one of two possible branches of the square root of $\bigl(\frac{a'}{a}\bigr)^2 - \frac{1}{a^2} 
= - \Lambda/3 \mpl^2$, and determines whether the membrane is placed on a latitude line excluding or including 
the equator on each $S^4$ section.   

These equations have important consequences. 
Plugging the second of Eqs. (\ref{metricjc}) into (\ref{diffroots}) shows that the RHS of (\ref{diffroots}) becomes
$\mp\frac{{\cal T}_{j} a}{4\mps}\Bigl(1 \mp {\tt Q}_{j} \Bigr)$ where, for fluxes with many units of charge, and using Eq. (\ref{totalcc}) as the definition
of the dependence of $\Lambda$ on fluxes and the charge $q$, such that $\Delta \Lambda = \frac{\beta}{\beta+\zeta}
\bigl(\alpha \mu^2 + \zeta ({\cal Q}-Q) \bigr) q$, 
\be
{\tt Q}_j = \frac{4\mps }{3 {\cal T}^2_{j}} \Delta \Lambda =  \frac{4\mps }{3 {\cal T}^2_{j}} \frac{\beta}{\beta+\zeta} 
\bigl(\alpha \mu^2 + \zeta ({\cal Q}-Q) \bigr) q ~~~ {\rm for} ~~~ j = Q ~{\rm or} ~ {\cal Q} \, .
\ee
To get these equations, we have used ${\cal Q} - Q \gg q$ and neglected ${\cal O}(q^2)$ terms; it is 
easy to restore these terms when fluxes involve only a few units of $q$. 

\begin{figure}[bth]
    \centering
    \includegraphics[width=7.5cm]{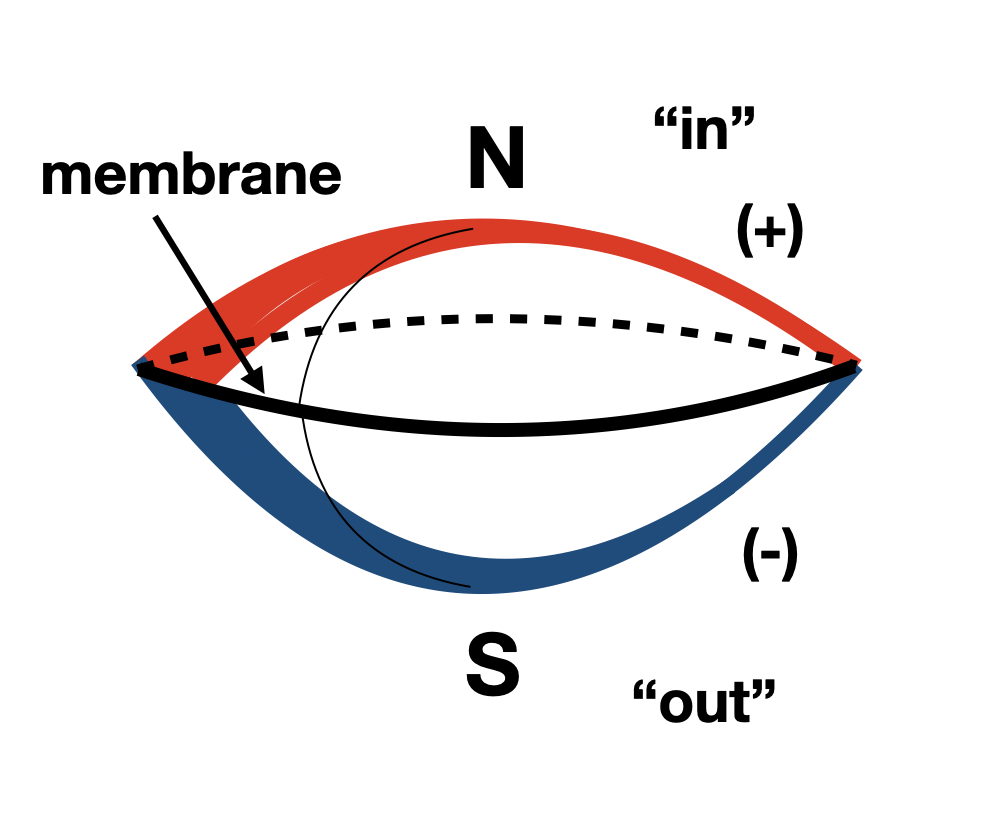}
    \caption{A $|{\tt Q}_j |<1$ instanton mediating $dS \rightarrow dS$.}
    \label{fig1}
\end{figure}

We can now see that when the flux contributions are small relative to $\alpha \mu^2/\zeta$,
\be
|{\cal Q}-Q| < \alpha \frac{\mu^2}{\zeta} \, , 
\label{qbounds}
\ee
the dominant contribution to the junction conditions, controlling the specifics of the instanton mediating $dS$ decay, comes from the 
{\it linear} term in the potential (\ref{abbottfin}), just like in \cite{Kaloper:2022oqv,Kaloper:2022utc,Kaloper:2022jpv,Kaloper:2023xfl}. 
When this happens, and we impose the conditions $|{\tt Q}_j |< 1$ for both $Q$ and ${\cal Q}$,
\be
{\tt Q}_j = \frac{4 \alpha \mu^2 \mps \, q}{3 {\cal T}^2_{j}}  \frac{\beta}{\beta+\zeta} < 1 \, ,
\label{greencond}
\ee
the quantity $\Bigl(1 \mp {\tt Q}_{j} \Bigr)$ is always positive. As we have explained in \cite{Kaloper:2022oqv,Kaloper:2022utc,Kaloper:2022jpv,Kaloper:2023xfl}, whenever this occurs the {\it only} instanton
which can mediate $dS \rightarrow dS$ decay is the instanton depicted in Fig. (\ref{fig1}).  Other $dS \rightarrow dS$ decay channels,  
which could in principle appear, are kinematically {\it forbidden}\footnote{There is {\it one more} 
instanton that mediates $dS \rightarrow AdS$ transition, which can happen only once 
\cite{Kaloper:2022oqv,Kaloper:2022utc,Kaloper:2022jpv,Kaloper:2023xfl}. 
Those regions are terminal sinks, which in the presence of matter collapse to form black 
holes, and can be ignored when it comes to accounting for the vacua of the theory.}. 

We can check the in-principle compatibility of (\ref{greencond}) with the Weak Gravity Conjecture. 
The inequalities of \cite{Arkani-Hamed:2006emk} amount to requiring that $\mpl \, q/{\cal T}_j > 1$ for fundamental membranes. 
 Since ${\tt Q}_j = \frac{4 \alpha \mu^2 \mpl}{3 {\cal T}_{j}}  \frac{\beta}{\beta+\zeta} \times \frac{ \mpl \, q}{{\cal T}_j}$,  
combining $\mpl \, q/{\cal T}_j > 1$ and  (\ref{greencond}) implies that  
\be
\frac{4 \alpha \mu^2 \mpl}{3 {\cal T}_{j}}  \frac{\beta}{\beta+\zeta} < \eta \, \frac{{\cal T}_j}{\mpl \, q} ~~~~ {\rm for} ~~~~ \eta < 1 \, , 
\label{wgcineq}
\ee
for both the Weak Gravity inequality and (\ref{greencond}) to 
hold simultaneously. For $\alpha \mu^2 < \mpl^2$ and sub-planckian tensions
${\cal T}_j$ this does not seem to be an insurmountable obstacle to careful model building.

In principle it is possible for the dynamics
to transition dynamically from a regime where $|{\cal Q}-Q| > \alpha \frac{\mu^2}{\zeta}$, where the instanton of Fig (\ref{fig1}) is forbidden,
and where a different effective theory is needed, 
to the regime where $|{\cal Q}-Q| < \alpha \frac{\mu^2}{\zeta}$, and the transitions can only occur via this instanton. This will happen naturally when the 
initial fluxes screening the cosmological constant are large, and as discharge proceeds and fluxes decrease, eventually the system will settle in
the regime where $|{\cal Q}-Q| < \alpha \frac{\mu^2}{\zeta}$. The important point is that the regime $|{\cal Q}-Q| < \alpha \frac{\mu^2}{\zeta}$ 
will naturally dominate as the total cosmological constant (\ref{totalcc}) decreases. 
The ``hard" cosmological constant $\Lambda_{\tt QFT}$ which needs to be cancelled by fluxes 
must be in this same regime, where the dominant screening in $\Lambda_{\tt total}$ comes from the linear terms in (\ref{totalcc}), 
for the discharge and screening to be fully reliable. Only then will  
the instanton of Fig. (\ref{fig1}) completely control the descent of $\Lambda_{\tt total}$ to the smallest reachable values. 
This does not happen without fine tuning in other common approaches to decaying the cosmological constant away by 
membrane discharges (see, e.g. \cite{Brown:1987dd,Brown:1988kg,Bousso:2000xa}), where due to the different structure 
of the theory other instantons control the decrease of $\Lambda_{\tt total}$, 
as detailed in \cite{Kaloper:2022oqv,Kaloper:2022utc,Kaloper:2022jpv,Kaloper:2023xfl}. 

Depending on the radius of the membrane at nucleation, there are 
two dynamical regimes when the instanton of Fig (\ref{fig1}) controls
the discharge. If we solve Eqs. (\ref{diffroots}) for the membrane radius $a^2$, we find  
\be
\frac{1}{a^2} = \frac{\Lambda_{out}}{3\mps} + \Bigl(\frac{{\cal T}_{j}}{4 \mps}\Bigr)^2 
\Bigr(1 -{\tt Q}_j \Bigr)^2 
= \frac{\Lambda_{in}}{3\mps} + \Bigl(\frac{{\cal T}_{j}}{4 \mps}\Bigr)^2 
\Bigr(1 +{\tt Q}_j \Bigr)^2 \, .
\label{radii}
\ee
Each term on the RHS is positive definite, and so as noted in \cite{Kaloper:2022oqv,Kaloper:2022utc,Kaloper:2022jpv}, the
generic options are controlled by the leading contribution to $a^2$. Initially, when the cosmological constant is large,  
$\Lambda_{out} > \frac{3}{16} \Bigl(\frac{{\cal T}_{j}}{\mpl}\Bigr)^2$, Eq. (\ref{radii}) gives $a^2 \propto (1-\frac{\Lambda_j a^2}{3 \mps})^{1/2} \ll 1$. At a quick glance, this suggests that in this limit the 
bounce action is $S_{\tt bounce} \simeq - \frac{12\pi^2 \mpl^4 \Delta \Lambda}{\Lambda_{out} \Lambda_{in}}$,
which is negative (but ${\cal O}(1)$ at he cutoff), 
and so the transitions with initially large cosmological constants are not very suppressed, and so can be fast. 
This bounce action resembles\footnote{The factor of $2$ difference in normalization can be expected 
since our transition occurs between two minima of the action, 
and not a minimum and an adjacent maximum. E.g. if we took the 
Hawking-Moss result, and applied it to $\Lambda_{out}$ and a value of $\Lambda$ half way between
$\Lambda_{out}$ and $\Lambda_{in}$ we'd get precisely (\ref{hawkmoss}).} 
the Hawking-Moss instanton transition rate \cite{Hawking:1981fz}.

However a more careful look shows that $\Lambda \rightarrow \infty$ and $\Lambda = 0$ are actually branch point singularities of $S_{\tt bounce}$ instead of poles, and one must do the limits more carefully. 
When $\Lambda \rightarrow \infty$ (or more precisely, in the limit ${\cal T}_j < {\cal M}_{\tt UV}^3$,
$\Lambda_{out} \simeq {\cal M}_{\tt UV}^4$, with ${\cal M}_{\tt UV} \sim \mpl$),
\be 
S_{\tt bounce} \rightarrow 0 \, ,
\label{fastbounce}
\ee
and so the decay rate is 
\be
\Gamma \rightarrow 1 \, ,
\label{hawkmoss}
\ee
meaning that the nucleations are almost completely unsuppressed. It is in this regime 
that most nucleations of bubbles occur. In fact since $S_{\tt bounce}$ increases toward 
$S_{\tt bounce} \simeq + \frac{12\pi^2 \mpl^4}{\Lambda_{out}}$ (see below) as 
the initial cosmological constant decreases, this rate 
disfavors the largest possible values of the cosmological constant, and favors the smallest ones, 
of as the terminal outcome of this stage, because the more curved backgrounds are
more unstable. 

Next, as $\Lambda$ decreases to $\Lambda \la \frac{3}{16} \Bigl(\frac{{\cal T}_{j}}{\mpl}\Bigr)^2$, 
the discharges are by the nucleation of small bubbles (relative to the horizon size), 
with the bounce action given by 
\cite{Kaloper:2022oqv,Kaloper:2022utc,Kaloper:2022jpv}
\be
S_{\tt bounce}  \simeq \frac{24\pi^2 \mpl^4}{\Lambda_{out}} 
\Bigl(1- \frac{8}{3} \frac{\mpl^2  \Lambda_{out}}{ {\cal T}_j^2} \Bigr)\, .
\label{familiars2}
\ee 
Now $S_{\tt bounce} > 0$ because $\Lambda < 3 \mpl^2  \Bigl(\frac{{\cal T}_j}{4 \mpl^2}\Bigr)^2$. Further, in this regime the nucleation 
rates become very small, since $\Lambda < 3 \mpl^2  \Bigl(\frac{{\cal T}_j}{4 \mpl^2}\Bigr)^2$ implies $\Lambda \ll \mpl^4$. This means,
this regime of $\Lambda$'s is ``braking" further evolution. Note, that the smaller the tensions ${\cal T}_j$,
the later this stage sets in, and the fast decays of the previous stage can 
populate the landscape of values of $\Lambda$ more efficiently. 

Reverse processes which increase the local value of the cosmological constant also occur. However they are more suppressed, as one can readily verify 
from using the formulas for the bounce actions we give, permuting the subscripts ``{\it out}" and ``{\it in}" and flipping the sign of 
$\Delta \Lambda$. Hence the dominant trend, when starting with some initial $\Lambda_{out} >0$, is the decrease of $\Lambda$. For large bounce actions,
which in light of our formulas are getting larger with the decrease of $\Lambda$, these processes slow down, 
implying the system is moving toward equilibrium. So in this regime 
we should be able to deduce the dynamical trends based on thermodynamic reasoning. 

The action (\ref{familiars2}) features a \underline{remarkable} property: it diverges as $\Lambda_{out} \rightarrow 0$!  
After a moment's thought this is not entirely surprising: as $\Lambda$ decreases the 
geometric entropy given by the de Sitter horizon area, ${\cal S}_{GH} \simeq \frac{24\pi^2 \mpl^4}{\Lambda_{out}} = \frac{A_{horizon}}{4G_N}$
grows, and the ensuing `chaos' takes over. The decay towards Minkowski then simply looks like the enforcement of
the $2^{\rm nd}$ law of thermodynamics. 
As a result the decay rate $\Gamma \sim e^{-S_{bounce}}$
has an essential singularity at $\Lambda_{out} \rightarrow 0$, where the rate vanishes \cite{Kaloper:2022oqv,Kaloper:2022utc,Kaloper:2022jpv}. 
Hence when $|{\tt Q}_j| < 1$, and the instanton of Fig. (\ref{fig1}) controls $dS \rightarrow dS$ transitions, 
small cosmological constants become very long lived. 
The closer the geometry gets to a locally Minkowski  space (with the cosmological constant that controls it being ``tossed" about vigorously in the
first stage), the more stable it becomes to discharges. If the 
cosmological constant ever ends up being zero, further discharges cease. This realizes the old idea that the distribution of terminal values of the
cosmological constant is controlled by the semiclassical Euclidean partition function of the theory \cite{Baum:1983iwr,Hawking:1984hk}. 
Some recent works \cite{Jacobson:2022jir,Jacobson:2022gmo} provide new support for this idea.  

Heuristically, in this regime we can glean why this is so by recalling the 
postulate that the ensemble average must be the time average of evolution,  
and get an estimate of the action which describes a sequence of nucleations. 
Since the decay processes governed by the instanton of Fig. (\ref{fig1}) slow down as $\Lambda$ decreases, in the ``braking" regime 
the semiclassical 
expression for the partition function $Z = \sum_{\Lambda}  \int \ldots  {\cal D} g \ldots \, e^{-{\cal S}_E}$
is dominated by our instantons, which split the sum into the sum over the instantons. 
To get a feel for it, we can estimate the individual terms in the sum by using the bounce action, as 
${\cal S}({\tt instanton}) = {\cal S}({\tt bounce}) + {\cal S}({\tt parent})$. In the case of multiple successive nucleations, this translates
to ${\cal S}({\tt instanton}) = \sum_n {\cal S}({\tt offspring}, n) + {\cal S}({\tt progenitor})$. By 
``offspring" we mean the geometric segments inside nested bubbles separated by the membranes, while 
the ``progenitor" is the primordial parent initiating the sequence - i.e., a patch of the initial de Sitter space. 

In this regime, when there is a single transition, Eq. (\ref{familiars2}) yields ${\cal S}({\tt instanton}) \simeq - 64 \pi^2 {\mpl^6}/{\cal T}_j^2$. 
By Eq. (\ref{radii}), when $\Lambda < 3 \mpl^2  \Bigl(\frac{{\cal T}_j}{4 \mpl^2}\Bigr)^2$, ${\cal T}_j^2 \sim 16 \mpl^4/a^2$,
and so the instanton action is proportional to the bubble area, ${\cal S}({\tt instanton}) \simeq - A_{bubble}/8G_N$. 
Subsequent transitions can occur as long as the interim cosmological constant is large enough (this requires a small charge $q$), 
with ${\cal S}({\tt instanton})$ increasing, approximately, by  
$\simeq - A_{bubble}/8G_N$ per step. This yields an estimate for a sequence of nucleations 
${\cal S}({\tt instanton}) \rightarrow - n A_{bubble}/8G_N$,  with $n$ the number of bubbles in the sequence, 
which is bounded by $ \frac{A_{horizon}}{4G_N} = - 24 \pi^2{\mpl^4}/{\Lambda}_{\tt terminal}$, 
for some terminal $\Lambda_{\tt terminal} \ga 0$. This 
implies that the sum $
Z \sim \sum \exp({24 \pi^2 \frac{\mpl^4}{\Lambda} + \ldots})$ favors the smallest achievable values of $\Lambda$ as long as the
flux discharge processes are mediated by the instanton of Fig. (\ref{fig1}). Hence 
the distribution of the terminal values of $\Lambda$ tends to be biased toward smallest values, peaking 
as $\sim \exp(24 \pi^2 \frac{\mpl^4}{\Lambda})$ at small $\Lambda$. 
Next we turn to the question of how small $\Lambda$ can get.  

\section{Landscape Painted with Fluxes}

The bound (\ref{qbounds}) which guarantees the dominance of the linear terms in (\ref{totalcc}) in the junction conditions (\ref{diffroots})
has extremely important implications for cosmological constant relaxation, as we will explain now. 
Since the fluxes $Q$ and ${\cal Q}$ are quantized in units of $q$ \cite{Teitelboim:1985ya,Teitelboim:1985yc}),  
$Q = Nq$ and ${\cal Q}= N' q$, Eq. (\ref{totalcc}) is 
\be
\Lambda_{\tt total} =  \Lambda_0 +\frac{\beta}{\beta+\zeta}  q \bigl(N' - N \bigr) \Bigl(\alpha \mu^2 + \frac{1}{2} \zeta q \bigl(N'- N \bigr) \Bigr) \, .
\label{totalccq}
\ee
Recalling Eq. (\ref{qbounds}), $|N'-N| < \alpha \mu^2/q\zeta$, to estimate how close $\Lambda_{\tt total}$ can get to zero we
can neglect the second term in the parenthesis in (\ref{totalccq}) to find that the closest $\Lambda_{\tt total}$ to zero is
\be
\Lambda_{\tt critical} \la \frac{\beta}{\beta+\zeta} \alpha \mu^2 q \, .
\label{lambdacrit}
\ee
So, if $\Lambda_{\tt critical}$ is smaller than $10^{-120} \mpl^4$ this can cancel the 
cosmological constant with satisfactory precision, but 
it will lead to cosmology with an empty universe problem \cite{Brown:1987dd,Brown:1988kg}, which we are striving to avoid. In 
\cite{Kaloper:2022oqv,Kaloper:2022utc,Kaloper:2022jpv,Kaloper:2023xfl} we proposed to resolve this problem by adding a second copy of the 
flux $\hat Q$, mimicking $Q$, with a charge $\hat q$ which is incommensurate with $q$: $\hat q = \omega q$ where $\omega$ is an irrational 
number. Transliterated here, that would imply that we add a second copy of the monodomy branch sector, spanned by fluxes $\hat Q, \hat {\cal Q}$,
quantized in the units of $\hat q$, and otherwise precisely replicating the $Q, {\cal Q}$ sector, such that 
\ba
\Lambda_{\tt total} &=&  \Lambda_0 +\frac{\beta}{\beta+\zeta}  q 
\bigl(N' - N \bigr) \Bigl(\alpha \mu^2 + \frac{1}{2} \zeta q \bigl(N'- N \bigr) \Bigr) \nonumber \\
&& ~~~ 
+ \frac{\hat \beta}{\hat \beta+\hat \zeta}  \hat q \bigl(\hat N' - \hat N \bigr) 
\Bigl(\hat\alpha \hat\mu^2 + \frac{1}{2} \hat\zeta \hat q \bigl(\hat N'- \hat N \bigr) \Bigr) \, .
\label{totalccq2}
\ea
Now, the two sectors could be incommensurable even when the charges $q, \hat q$ are not mutually irrational. All it takes 
is for the mass scales $\mu, \hat \mu$ to be generic, as induced by some gauge theory strong dynamics, and the irrational ratios would appear
immediately. 

However, the problem which this has to face is the bound (\ref{qbounds}) for $Q, {\cal Q}$ 
and its equivalent for the duplicates  $\hat Q, \hat {\cal Q}$, which imply  $|N'-N| < \alpha \mu^2 /q\zeta$ 
and  $|\hat N'-\hat N| < \hat \alpha \hat \mu^2 /\hat q \hat \zeta$, and as a result,
\be
|N| \la \frac{ \alpha \mu^2}{q\zeta} \, ,  ~~~~~ |N'| \la \frac{ \alpha \mu^2 }{q\zeta} \, , 
~~~~~ |\hat N| \la \frac{\hat \alpha \hat \mu^2 }{\hat q \hat \zeta} \, , ~~~~~ |\hat N'| \la \frac{\hat \alpha \hat \mu^2}{\hat q \hat \zeta} \, .
\label{nbox}
\ee
This means that the bound (\ref{qbounds}) ``boxes in" the integer units of fluxes inside a 4-`cuboid', of sides given by the RHS
in inequalities (\ref{nbox}). The integers $N, N', \hat N, \hat N'$ cannot be arbitrarily large. Hence the number of possible values of the
cosmological constant $\Lambda_{\tt total}$ that fall into this box is {\it finite}, albeit it can be huge. 
Since finite sets are not dense in the set of real numbers, the arguments relying on the mutual irrationality of fluxes 
approximating any real number arbitrarily closely are obstructed, since the integers $N, N', \hat N, \hat N'$
are bounded. 

In our case, if we were to 
allow the integers $N, N', \hat N, \hat N'$ to be arbitrarily large, in addition to violating the bounds (\ref{qbounds}), (\ref{nbox}), we would
de facto allow the axions $\phi, \hat \phi$ (which are integrated out at low energies) to acquire arbitrarily large field excursions,
effectively decompactifying them. As a consequence the discrete shift symmetries could become global continuous symmetries. 
Based on the lore that quantum gravity prohibits this \cite{Banks:1991mb,Banks:2010zn}, we infer that at least in this case the 
bounds (\ref{qbounds}), (\ref{nbox}) must not be violated, and so the irrational superpositions of a couple of fluxes 
alone might not span a sufficiently dense set of values of the cosmological constants (or, equivalently, a sufficiently 
large set of low energy vacua). 

\begin{figure}[bth]
    \centering
    \hskip.4cm
    \includegraphics[width=9.5cm]{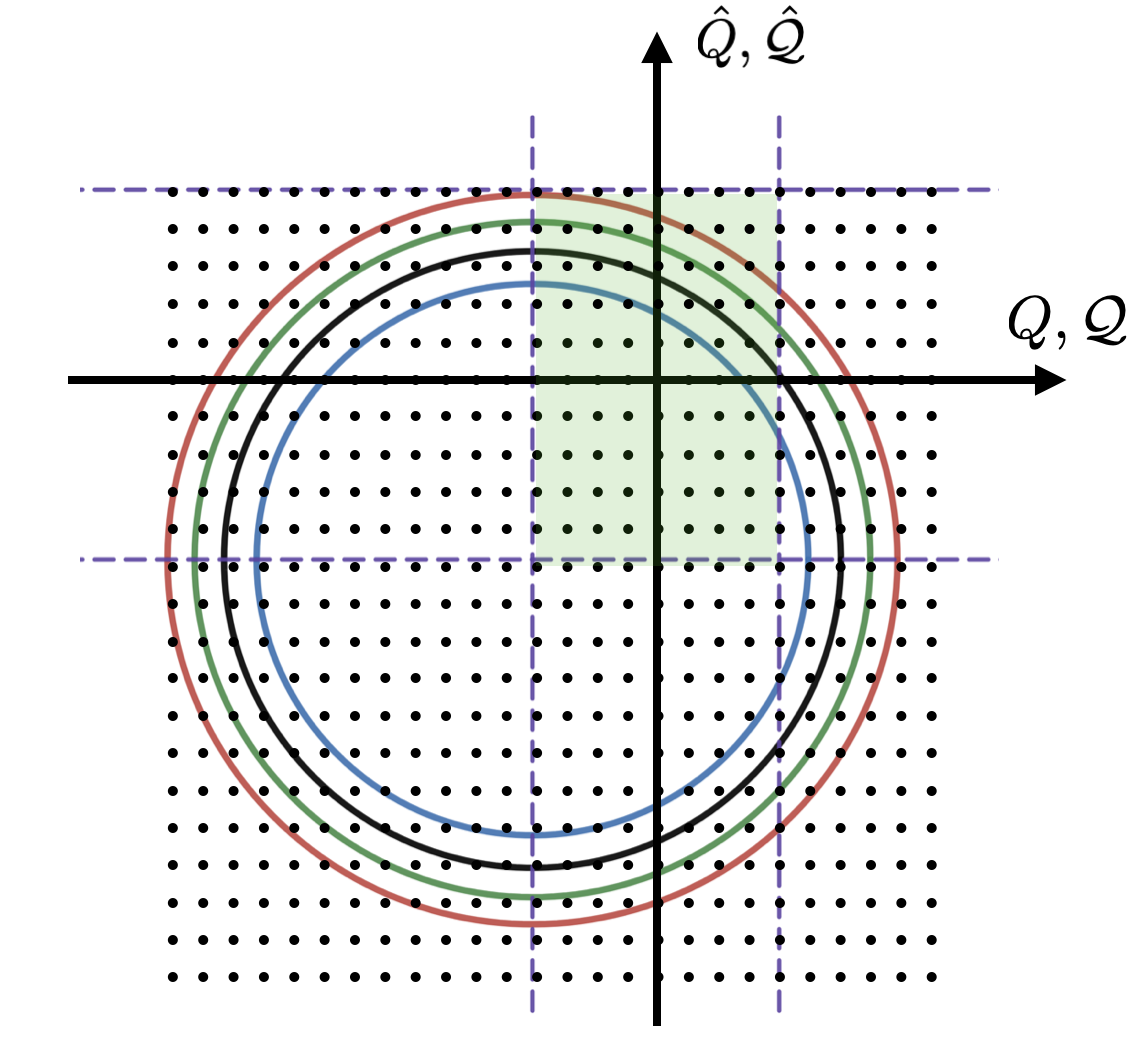}
    \caption{The phase space of values of $\Lambda_{\tt total}$ in the charge lattice (with only one representative of each pair of discrete variables shown). 
    The dashed vertical and horizontal lines are the boundaries of the region of phase space where the charges satisfy the bounds (\ref{qbounds}), 
    (\ref{nbox}) and their copies, which is
    lightly shaded. Inside this region, both the linear terms in fluxes dominate the junction conditions, and the grid points can be
    close to the surface of the fixed cosmological constant.}
    \label{counts}
\end{figure}

A clue to resolving this problem comes by considering (\ref{totalccq2}) as a system of hypersurfaces in the lattice spanned by 
the integers $N, N', \hat N, \hat N'$. First, it is convenient to complete the squares in  (\ref{totalccq2}) and rewrite this equation as
\ba
&& \frac{\beta \zeta}{\beta+\zeta} 
 \Bigl(q\Delta N + \frac{\alpha \mu^2}{\zeta} \Bigr)^2 +  \frac{\hat \beta \hat\zeta}{\hat\beta+\hat\zeta}
\Bigl(\hat q\Delta \hat N + \frac{\hat\alpha \hat\mu^2}{\hat\zeta} \Bigr)^2 = 2 \Lambda_{\tt eff} \, , \nonumber \\
&& ~~~~~ 2 \Lambda_{\tt eff}  = 2(\Lambda_{\tt total} -  \Lambda_0) 
+ \frac{\beta \alpha^2 \mu^4}{(\beta+\zeta)\zeta} + 
\frac{\hat \beta \hat \alpha^2 \hat\mu^4}{(\hat\beta+\hat \zeta)\hat \zeta} \, ,
\label{spheres}
\ea
where $\Delta N = N - N'$ and $\Delta \hat N = \hat N - \hat N'$. These hypersurfaces are circles when 
projected to the diagonal flat hyperplanes in the $4D$ space of pairs of charges $(N, N'), (\hat N, \hat N')$. We plot them for a fixed 
$\Lambda_{\tt total}$ in Fig (\ref{counts}).  

What this figure shows is, that any initial value of the cosmological constant $\Lambda_{\tt eff}$ can 
cascade down to near zero by a sequence of membrane 
nucleations that discharge $Q, {\cal Q}, \hat Q, \hat {\cal Q}$, ending up at a distance from zero given by the distance of the closest lattice point
in the grid depicted in Fig. (\ref{counts}) to the initial surface $\Lambda_{\tt eff}$. All we need to do is to make sure that the grid in Fig. (\ref{counts}) is
sufficiently refined to ensure that the initial distance is $\la 10^{-120} \mpl^4$ and the problem should be solved. This means, we adopt the criterion 
of charge grid density of \cite{Bousso:2000xa} to our case. The only difference is that in our case the surfaces of constant $\Lambda$ are not centered 
at the origin, due to the linear flux terms which shift the centers away, and that the sections which fall into the allowed shaded region
are not complete circles, but instead their arcs. However since the centers are shifted by the distance comparable to the side of the shaded
box of allowed values of fluxes (\ref{qbounds}), these arcs span ${\cal O}(1)$ of the full circle, and this restriction is 
of little practical consequence. 
There will be points near the arcs, and the cascade of flux discharges does not have to ever step outside the allowed shaded box. 

To guarantee sufficient grid refinement, we will need to add more copies of the flux monodromies, and determine their number by requiring that there
is always a grid point near a surface of a given $\Lambda$ inside the allowed region. With $J$ copies of $N, N'$, Eq. (\ref{spheres}) generalizes to
\be
\sum_{j=1}^J a_j \Bigl(q_j \Delta N_j +  b_j\Bigr)^2 = 2(\Lambda_{\tt total} -  \Lambda_0) 
+ \sum_{j=1}^J a_j b_j^2 \, ,  ~~~~~~ a_j =  \frac{\beta_j}{\beta_j+\zeta_j}\, ,  ~~~~~~ b_j = \frac{\alpha_j \mu_j^2}{\zeta_j} \, . 
\label{spheresJ}
\ee
To figure out how closely the grid points will be to this surface, we need to compute their number, given that their density is one per cell,
between two concentric arcs in Fig. (\ref{counts}) in the shaded region. From Eq. (\ref{spheresJ}) we deduce that
$\delta \Lambda = \sum_{j=1}^J a_j q_j \Bigl(q_j \Delta N_j +  b_j\Bigr) \delta \Delta N_j \simeq \sqrt{2 \Lambda_{\tt total}} \sqrt{a_j} q \delta N$, 
because by naturalness all the terms on RHS of (\ref{spheresJ}) are comparable. Next, the volume between two
concentric shells is given by the shell area times $\sqrt{a} q\delta N$, which yields
$d{\cal V} = \Omega_{J-1} \rho^{J-1} \sqrt{a_j} q \delta N$, where $\rho^2 = \sum_{j=1}^J a_j \Bigl(q_j \Delta N_j +  b_j\Bigr)^2$, and  
$\Omega_{J-1}$ is the solid angle measuring the extent of the arcs fitting in the shaded regions in Fig. (\ref{counts}).  It is 
by our construction a fraction of the full solid angle in $J$ dimensions. The coefficients
$\sqrt{a_j} q_j$ play the role of Lam\'e's coefficients of vector calculus. Since generally $a$'s and $q$'s are different, the degeneracy of these
states is $D=2$ since there is a pair of $N,N'$ for each direction, which are  clearly degenerate. However, there is also an enhancement for each
direction, since any number $N+N'$ can be realized as multiple sums of two integers. There is roughly $N$ possibilities for each directions. 

Putting it all together, the number of points inside the arc shell is
\be
\delta P \simeq \frac{1}{\bigl(\prod_j^{J} \sqrt{a_j} q_j \bigr)}  
\bigl(\prod_j^{J} \frac{N_j}{2} \bigr) \Omega_{J-1} \bigl(2 \Lambda_{\tt total}\bigr)^{\frac{J}{2}-1} \delta \Lambda \, , 
\ee
where the first factor converts the density of integers into the density of fluxes. Now we require that there is at least one point inside the shell,
$\delta P \ga 1$, and that the thickness of the shell corresponds to the observationally acceptable gap for cosmological constant \cite{Bousso:2000xa}.
This yields
\be
\bigl(\prod_j^{J} \frac{N_j}{2} \bigr) \Omega_{J-1} \bigl(2 \Lambda_{\tt total}\bigr)^{\frac{J}{2}-1}  \mpl^4 \ga 10^{120} \bigl(\prod_j^{J} \sqrt{a_j} q_j \bigr) \, ,
\label{density}
\ee
as the formulation of the task which a model builder must achieve to produce a viable model.
To get the idea of the scales involved, we can consider some simple limits. Let's imagine that all the field space directions are
comparable, which means that parameters for each `direction' in the monodromy space are about the same. The 
phase space solid angle is given by a fraction $\Theta$ of the area of a unit $J$-dimensional sphere, 
so $\Omega_{J-1} = \frac{2\pi^{J/2}}{\Gamma(J/2)} \Theta $.  Finally, assuming naturalness, 
let $\Lambda_{\tt total}$ and $\Lambda_0$ be comparable with the
cutoff contribution $\sum_{j} a_j b_j^2 \sim J \alpha^2 \mu^4/\zeta^2$, and $a_j \simeq 1$ while $\zeta \ll 1$ to produce a
long regime of linear flux dominance. Using Eq. (\ref{nbox}) for the estimate of $N_j$, and substituting all these estimates into (\ref{density}),
we finally obtain
\be
\Theta \frac{\pi^{J/2}}{2\Gamma(\frac{J}{2})} \bigl(\frac{J}{4}\bigr)^{\frac{J}{2}-1} \bigl(\frac{\alpha \mu^2}{\zeta q} \bigr)^{2{J}-{2}}  \ga  
10^{120} \frac{q^2}{\mpl^4} \, .
\label{isodensity}
\ee
If we were to take all the dimensional scales to be ${\cal O}(0.01 \mpl)$, we'd be able to obtain the right refinement of the monodromy grid
with taking $\zeta \sim 0.01$ and using at least about 30 pairs of fluxes $Q, {\cal Q}$. 
The monodromy flux landscape that would ensue would be fine enough
to ensure satisfactory cancellation of the cosmological constant values which are originally near the cutoff $\simeq \mpl^4$. Indeed, this means that 
the ``flux box" defined by the bounds (\ref{nbox}) would contain about $[(1/\zeta)^2]^J \sim [(100)^2]^{30} \sim 10^{120}$ vacua. 

From Eq. (\ref{isodensity}), we see that the main scale dependence comes from the ratio $q^2/\mpl^4$ on the RHS of the equation,
since given the quantization condition (\ref{quant}),  the other scale ratios are 
$\frac{\mu^2}{q} = \frac{\mu}{2\pi f} = \frac{q}{4\pi^2 f^2}$. The previous example, and in particular the interpretation of the result 
in terms of the number of vacua inside the box (\ref{nbox}), points to how to achieve high refinement of the flux grid. We want to
find many vacua in the box (\ref{nbox}), which implies that the axion field range $\Delta \phi = 2\pi f$ should be smaller
than the slope parameter $\mu$. This is consistent with having efficient axion trapping by the potential wells. So if we take
$\alpha \mu/2\pi f \sim 100$, in addition to taking $\zeta \sim 0.01$ to extend the domain of dominance of the linear flux terms, and for
example drop $\sqrt{q}$ to $10^{-8} \mpl \sim 10^{10} {\rm GeV}$, we can satisfy Eq. (\ref{isodensity}) with 
at least $J \sim 12$ pairs of
degenerate fluxes $Q, {\cal Q}$. In this case the ``flux box" would contain $\sim 10^{90}$ vacua.
Obviously, a more precise determination of scales would require a more precise construction of the vacua of the theory, e.g. along the
lines of the papers \cite{Higaki:2014mwa,Bachlechner:2015gwa,Bachlechner:2017zpb,Bachlechner:2017hsj}. Note, that in the more general
case we would add into the fray not only potentials with equal periods for the pairs 
$Q, {\cal Q}$, but with general mutually rational periods, and also 
generic phase differences. This would produce an even greater diversity of vacua than considered here. In fact, it has already been
shown that in the case of the multi-dimensional cosine potentials, the number of vacua can be very large, but with  accumulation near the smaller 
values of total cosmological constant \cite{Masoumi:2016eqo}. Adding the mass term, which arises in flux monodromy, to this model 
spreads out the vacua further when only one monodromy branch is retained. 
We expect that with the addition of full monodromy structure
the overall picture will remain, albeit the density of vacua could be higher toward lower values 
of the cosmological constant. This could aid the dynamics we discussed further. We hope 
to return to this very interesting issue in the future. 

The model ingredients which we have touched upon here point to the following scenario. We can imagine a theory where supersymmetry (SUSY)
is broken at some scale $(\alpha \mu^2/\zeta)^{1/2}$, triggering chiral symmetry breaking in some strongly coupled sectors. As a result, the 
axion in those sectors becomes massive and decouples, while the theory has many vacua which can evolve into each other
by membrane discharges. At a scale above SUSY breaking, SUSY itself cancels the cosmological constant. Below SUSY breaking, the
membrane discharges do it, searching statistically for the near-Minkowski vacuum when the linear flux terms dominate below SUSY breaking. 
Since we restricted the range of fluxes to be within the box of the size given by the field ranges that guarantee the linear flux term dominance, 
the low energy spectrum of cosmological constants has a finite gap, and the evolution favors the value of the cosmological constant at a value given 
by this gap. 

Thus with finding the models where the parameters saturate the inequality (\ref{isodensity}) we can accommodate the terminal value of the 
cosmological constant to be precisely $10^{-120} \mpl^4$. Given the attractor behavior of the theory, because the transitions are mediated
by the instanton of Fig (\ref{fig1}), evolving toward this small terminal value 
does not require anthropic reasoning. During the initial stage the initial cosmological constant
is discharged faster, while in the second, ``braking" stage, when the cosmological constant is small, the 
discharge rate is exponentially slow. We note that it is conceivable that in some other boroughs of this monodromy 
landscape the linear flux dominance is violated, and the distribution of 
the cosmological constant values is flatter. However
as long as the regions we describe exist, they may win, thanks to the attractor. 

\section{Cosmological Connections}

By the construction of the model here, the cosmological constant evolution occurs in discrete steps, 
and the membrane charges are not minuscule. Therefore the 
value of the cosmological constant just before the last membrane is nucleated, in whose interior the residual $\Lambda$ is tiny, is in fact large.
It is plausible that when the last nucleation occurs, before the terminal value of $\Lambda$ is eventually reached, the spacetime can be 
dominated by a transient regime of inflation. After this stage of inflation ends, the universe can reheat and become repopulated, 
unlike what happens in Abbott's model with a smooth variation of the adjustment field. Such evolution also would not completely erase 
the future of the information about the ancestry of the final near-Minkowski space. Since discrete adjustments happen locally, at large scales 
the universe is composed of the regions separated from each other by highly curved spacetime, with large 
cosmological constant values, which are still bubbling.

It would be interesting to connect our mechanism with inflation in a more precise way. An avenue, which we think is promising, given the
``monodromy genetics" of our mechanism, is to embed it in general theory of monodromy inflation. In the first pass, it might suffice to simply 
flatten out the potential of one of the axions which we introduced above, and study how to embed the quadratic flux monodromy model of
\cite{Kaloper:2008qs,Kaloper:2008fb,Kaloper:2011jz}, despite the fact that this specific model does not fit the cosmological observations 
anymore. That, by itself, is not a problem since there remain phenomenologically completely viable variants of monodromy inflation, which 
continue to fit the observations perfectly, such as the extremely flattened models of \cite{Dubovsky:2011tu,Nomura:2017ehb}, 
or unwinding inflation of \cite{DAmico:2012wal,DAmico:2012khf}.
A successful blend of our mechanism with monodromy inflation would provide a complete model of 
relaxing $\Lambda$ without anthropic reasoning, which would also 
produce a number of observationally testable predictions. 

Another very interesting cosmological question concerns phase transitions in the early universe. In previous work 
\cite{Kaloper:2022oqv,Kaloper:2022utc,Kaloper:2022jpv,Kaloper:2023xfl} we have raised the question about phase transitions that may occur
{\it after} the membrane discharge processes effectively decouple because the background cosmological constant becomes very 
small and quantum evolution slows down. In this case one can worry if a phase transition in the matter sector changes the QFT vacuum energy,
since it would generically decrease $\Lambda_{\tt background}$ to a large negative value. Such regions 
would stop expanding and recollapse, presumably 
forming a large black hole. 

However a possible resolution of this issue is that in the very early universe, when the background cosmological
constant is large and positive, the symmetries which are broken by these phase transitions are {\it already} broken. 
Namely consider the example of a QFT with spontaneous symmetry breaking controlled by a field with a mass $m$. 
Let's take\footnote{For light fields the corrections coming from de Sitter geometry must be accounted for. 
However note that for e.g. electroweak sector, with $m_{\tt Higgs} \sim 100 \, {\rm GeV}$,
$\Lambda_{\tt total}$ can be as high as $(10^{10} \, {\rm GeV})^4$, and our argument should still apply with little modification. This precisely fits our
second example of monodromy landscape from the previous section.} 
$m^2 > \Lambda_{\tt total}/(3\mps) = H_{\tt total}^2$, so that we can neglect the de Sitter corrections to the mass $m$. Since the background
cosmological constant quickly supercools the universe, the field will roll to the true vacuum instantaneously, changing the total cosmological 
constant to $\Lambda_{\tt total} - \Lambda_{\tt latent}$, where $\Lambda_{\tt latent}$ is the latent heat of the phase transition. 
In all such bubbles, therefore, the contribution from the phase transitions will be already included and accounted for
in the matter sector vacuum energy, which is being screened and discharged by fluxes. The discharges will bring 
$\Lambda_{\tt total} - \Lambda_{\tt latent}$ close to zero, not $\Lambda_{\tt total}$. Afterwards, this symmetry may be restored or not, 
depending on the reheating, but since the early de Sitter has already ``prepared" the initial state of the universe, the vacuum 
energy of the terminal state will be already `prearranged' and the phase transitions later on, if they reoccur, 
will not result into decay to $\Lambda_{\tt terminal} < 0$, but will end up 
with a universe with $\Lambda_{\tt terminal} \ga 0$, like our own. It wold be very interesting to 
study this in more detail, to see how to properly account for the vacuum energy changes induced by electroweak and QCD phase transitions. 

\section{Summary}

What is a simple intuitive description of the mechanism for cosmological constant discharge which we have been developing here and in 
\cite{Kaloper:2022oqv,Kaloper:2022utc,Kaloper:2022jpv,Kaloper:2023xfl}? Recall first that the cosmological constant is measured by 
the magnetic fluxes quantized in the units of charge $q$, which are space filling. To visualize this, let's suppress one spatial dimension, 
and think of how this mechanism appears in a $2+1$-dimensional universe. Imagine a stack of $N$ ``rubber sheets" with tension, 
which stretch from one end of the universe to the other. Each sheet carries a unit $q$ of flux, equal to the membrane charge, 
and so their total number measures the cosmological constant over a `zero' level set by 
some $\Lambda_{\tt QFT}$. 
\begin{figure}[bth]
    \centering
    \includegraphics[width=12.37cm]{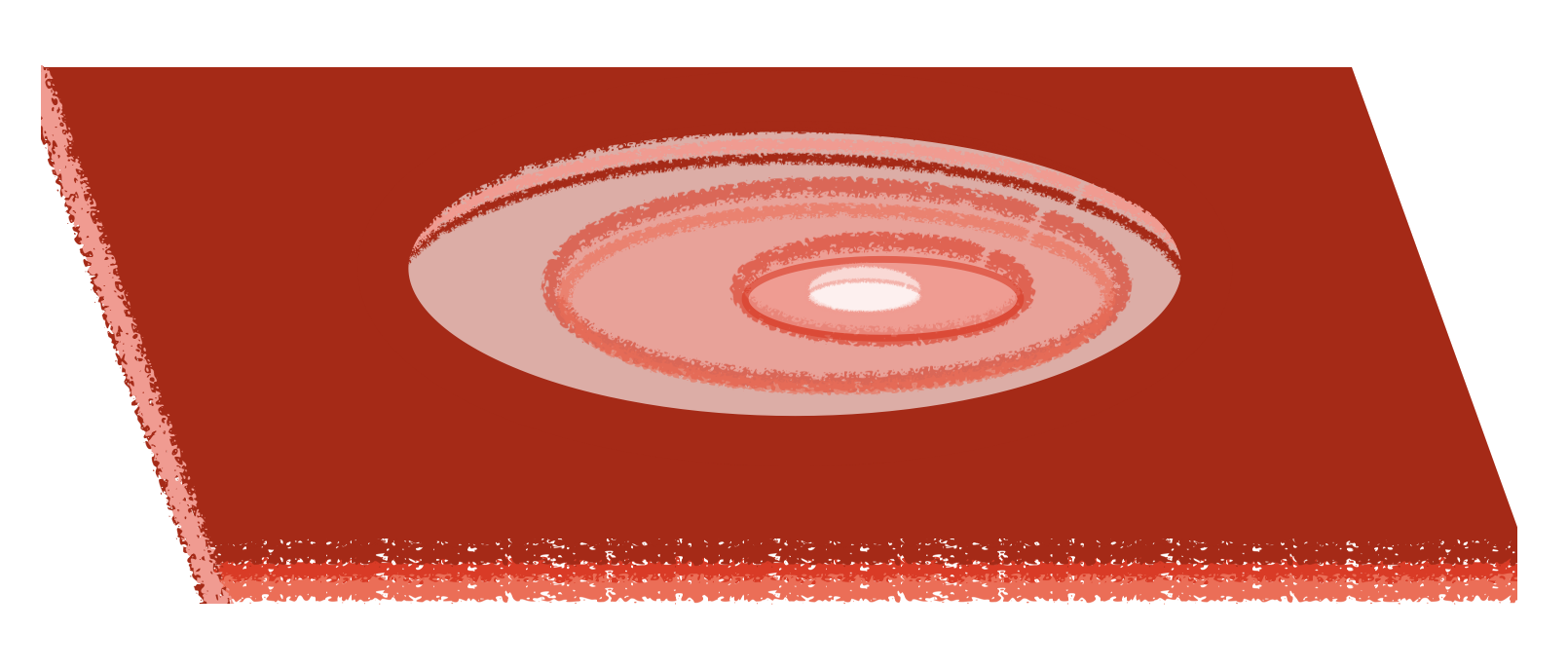}
    \caption{A stack of fluxes (red sheets) with holes, which represent the bubbles of successively smaller cosmological constants.}
    \label{sheets}
\end{figure}

Now imagine piercing a very small hole in the top sheet. Since the sheet has tension, the hole will stretch and expand
as a perfect circle. The interior will have $N-1$ sheets which are not pierced yet, and it represents the interior of a bubble with a cosmological
constant reduced by $q$ relative to the exterior. The expanding edge of the rupture is the membrane, which is expanding out to infinity. Then 
pierce the next sheet, and the next one. Each time a hole is pierced, a new bubble forms, and a new circular edge stretches out, with further 
reduction of $\Lambda$ inside. Then finally declare that the rate at which piercings occur is controlled by the number of sheets above the level of
$\Lambda_{\tt QFT}$ which are not pierced yet. The fewer there are, the rate decreases dramatically. Eventually, the rate of new tears comes to a full stop. 
This behavior reminds of how stimulated
emission occurs in quantum systems. This is depicted in Fig. (\ref{sheets}). 

To approach the background value of $\Lambda_{\tt QFT}$ very closely, just one type of sheets will not do if the 
charge $q$ is not tiny, and $\Lambda_{\tt QFT}$ is not fine tuned. This is why we need many species of fluxes, 
whose joint evolution and refined charge grid, which is very dense, allow the system to come very close 
to $\Lambda_{\tt QFT}$. In the language of our analogy, we need a very colorful system of ``rubber sheets" 
to evade fine tuning.

Here we have outlined how to realize such dynamics using axion monodromies which involve very heavy axions but contain flattened flux contributions.
The flux-dependent potentials span a monodromy landscape in field theory, and branch changes by membrane discharge will 
discharge the cosmological constant. When we constrain the total flux variations to reflect the limits on the field ranges, we find that the 
terminal values of the cosmological constant, favored by quantum attractor evolution, are tiny but finite. They can explain the current 
observations of dark energy scale. It will be interesting to extend the theory by incorporating slow roll inflation, analyze aspects of
field theory phase transitions and provide specific constructions of the monodromy landscapes. 

\vskip.5cm

{\bf Acknowledgments}: 
We thank G. D'Amico, 
A. Lawrence, J. Terning and A. Westphal for valuable comments and discussions. NK is supported in part by the DOE Grant DE-SC0009999.

\end{document}